\documentclass[conference]{IEEEtran}

\IEEEoverridecommandlockouts
\usepackage[pdftex]{graphicx}
\usepackage[cmex10]{amsmath,empheq}
\usepackage{cite}
\usepackage{url}
\usepackage{color}
\usepackage{microtype}
\usepackage{amssymb}
\usepackage{relsize}
\usepackage{lipsum}
\usepackage{epstopdf}
\usepackage{enumitem}
\usepackage{bm}
\usepackage{nicefrac}
\usepackage[percent]{overpic}
\usepackage{multirow}
\usepackage{tikz}
\usepackage{pgfplots}
\usepackage{flushend}
\usepackage{fancyhdr}

\fancypagestyle{firstpage}
{
  \fancyhf{}
  
  \fancyfoot[C]{\footnotesize \copyright 2023 IEEE. Personal use of this material is permitted. Permission from IEEE must be obtained for all other uses, in any current or future media, including reprinting/republishing this material for advertising or promotional purposes, creating new collective works, for resale or redistribution to servers or lists, or reuse of any copyrighted component of this work in other works.}
}

\usetikzlibrary{calc}
\usepgfplotslibrary{groupplots}
\pgfplotsset{compat=1.18}

\usepgfplotslibrary{fillbetween}

\begin{document}

\title{Deep Learning-Based Pilotless Spatial Multiplexing}

\newcommand{\todo}[1]{}
\renewcommand{\todo}[1]{{\color{red}{#1}}\PackageWarning{TODO:}{#1!}}
\newcommand{\changed}[1]{{\color{blue}{#1}}}

\author{
\IEEEauthorblockN{Dani Korpi, Mikko Honkala, and Janne M.J. Huttunen}
\IEEEauthorblockA{\textit{Nokia Bell Labs}\\
\textit{Espoo, Finland\vspace{-3mm}}}
}

\maketitle

\begin{abstract}
This paper investigates the feasibility of machine learning (ML)-based pilotless spatial multiplexing in multiple-input and multiple-output (MIMO) communication systems. Especially, it is shown that by training the transmitter and receiver jointly, the transmitter can learn such constellation shapes for the spatial streams which facilitate completely blind separation and detection by the simultaneously learned receiver. To the best of our knowledge, this is the first time ML-based spatial multiplexing without channel estimation pilots is demonstrated. The results show that the learned pilotless scheme can outperform a conventional pilot-based system by as much as 15--20\% in terms of spectral efficiency, depending on the modulation order and signal-to-noise ratio.
\end{abstract}
\thispagestyle{firstpage}

\section{Introduction}

Using artificial intelligence (AI)  for designing and learning more efficient physical layer solutions is an emerging concept in the field of wireless communications \cite{Honkala21}. It is even envisioned that future 6G networks could rely on a fully AI-native air interface where deep learning is an integral part of the physical layer of the radio system~\cite{Hoydis21a}. In the most extreme scenario, this could mean that both the transmit waveform and the receiver algorithm are learned from data. Such an approach is typically referred to as end-to-end learning.

In this work, we propose one approach for enhancing the physical layer with deep learning techniques, showing clear throughput gains. In particular, we train multiple-input and multiple-output (MIMO) transmitter and receiver end-to-end to communicate without any channel estimation pilots. This means that the link can perform spatial multiplexing without wasting any resources for pilot overhead, thereby improving the spectral efficiency. In practice, this is achieved by considering the transmitter and receiver as part of a single model, connected by a differentiable wireless channel. Such a system can be learned in a supervised manner by treating the channel and any required conventional functions as fixed layers, which are not modified during the training.

Machine learning (ML)-based pilotless systems have earlier been shown to be feasible for single-input and single-output (SISO) systems, where no spatial multiplexing is performed \cite{aoudia20}. As for MIMO communications, the work in \cite{He20} shows a model-based deep learning approach for training an accurate MIMO detector. In \cite{Korpi21}, on the other hand, a fully learned MIMO receiver is proposed, and it is shown to outperform conventional linear minimum mean square error (LMMSE) receivers. Moreover, \cite{pratik20} also proposes a fully learned ML-based MIMO receiver, which is shown to be able to handle a varying number of users while achieving state-of-the-art demodulation accuracy. Additional ML-based MIMO detectors are proposed in \cite{Huang18,Liao20}. However, to the best of our knowledge, there are no prior works considering end-to-end learning of MIMO links to achieve pilotless transmissions.

\section{System Model and Proposed Scheme}

\begin{figure*}[t!]
	\centering
	\includegraphics[width=\textwidth,trim={0cm 14.5cm 5.5cm 0cm},clip]{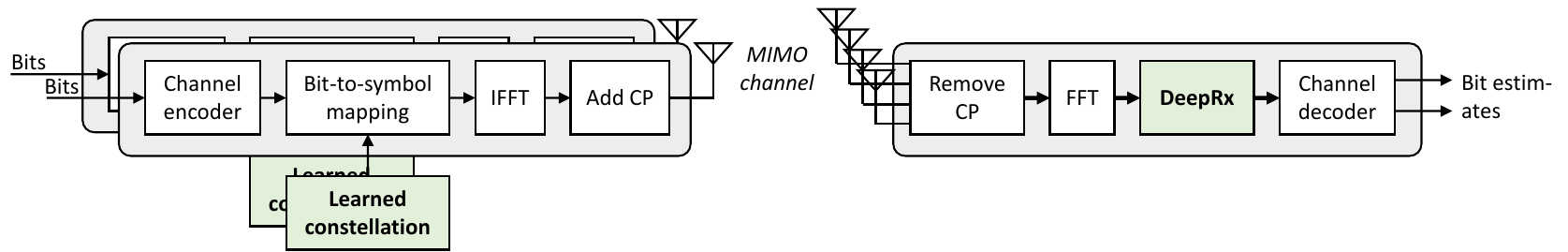}
	\caption{The considered system model for the ML-based pilotless scheme.}
	\label{fig:system_model}
\end{figure*}

A high-level depiction of the considered system model is presented in Fig.~\ref{fig:system_model}. In particular, we consider an orthogonal frequency-division multiplexing (OFDM) system, where a learned constellation shape is utilized to facilitate fully blind MIMO detection at the receiver side using a DeepRx-type convolutional neural network (CNN)-based receiver \cite{Honkala21}. The detailed architecture of the considered ML-based receiver follows the MIMO DeepRx presented in \cite{Korpi21}, using the fully learned multiplicative transformation. In particular, the ML receiver takes in the Fourier transformed received MIMO signal and outputs the log-likelihood ratios (LLRs) for each received spatial stream. With this, the input signal to the receiver can be expressed as:
\begin{align}
\mathbf{y}_{ij} = \mathbf{H}_{ij} \mathbf{x}_{ij} + \mathbf{n}_{ij}, \label{eq:y_ij}
\end{align}
where $\mathbf{H}_{ij} \in \mathcal{C}^{N_R \times N_T}$ is the MIMO channel matrix for the $i$th subcarrier and $j$th OFDM symbol, $N_R$ is the number of receive antennas, $N_T$ is the number of transmitted spatial streams, $\mathbf{x}_{ij} \in \mathcal{C}^{N_T \times 1}$ is the transmitted symbol vector, and $\mathbf{n}_{ij} \in \mathcal{C}^{N_R \times 1}$ is the additive noise signal.

\subsection{Learning MIMO Constellations}

In order to achieve pilotless spatial multiplexing, the transmit constellations are parametrized and learned jointly with DeepRx. This means that, in the transmitter, the elements of $\mathbf{x}_{ij}$ are selected from learned constellation points: 
\begin{align}
\mathbf{x}_{ij} \in \mathcal{M}(\bm{\theta})= \{\mathbf{x}\in\mathcal{C}^{2^{Q_m} \times N_T}:\ \mathbf{x}=N_{\theta}(\mathbf{q}_\textrm{QAM})\},
\end{align}
where $Q_m$ is the modulation order. In other words, the learned constellations are given by a neural network $N_\theta$ transforming the standard QAM-constellation points $\mathbf{q}_\textrm{QAM}$ to $N_T$ different variations. Indeed, pilotless spatial multiplexing is facilitated by learning separate constellations for each MIMO layer.

The more detailed procedure for generating the transformed constellations involves layer-wise neural network models, which are used as follows to provide the per-layer learned constellation:
\begin{align}
N_{\bm{\theta},l}(\mathbf{q}_\textrm{QAM}) = \sum_{c=1}^{C} W_{\bm{\theta},l,c}(\mathbf{q}_\textrm{QAM}) Q_{\bm{\theta},c}(\mathbf{q}_\textrm{QAM})\in\mathcal{C}^{2^{Q_m} },
\end{align}
where $C$ is a hyperparameter identifying the number of independent learned transformations of the regular QAM constellation, $Q_{\bm{\theta},c}(\mathbf{q}_\textrm{QAM})$ is a neural network producing the $c$th transformation of the QAM constellation, and $W_{\bm{\theta},l,c}(\mathbf{q}_\textrm{QAM})$ is a neural network which provides the weight factors of the learned transformations for each layer. In other words, the learned constellations for each layer are formed as a linear combination of learned transformations of the initial QAM constellation. Moreover, non-zero mean is removed from the learned constellations and they are normalized to have unit energy, which ensures that the ML-based scheme has the same overall transmit energy as a conventional OFDM transmitter (assuming uniformly distributed transmit bits).

The neural network for producing the transformation, denoted by $Q_{\bm{\theta},c}(\mathbf{q}_\textrm{QAM})$, consists of four fully-connected hidden layers with hyperbolic tangent activations and 16--32 neurons per layer, while the output layer has a linear activation function. The input of the neural network consists of the amplitude and angle of the original QAM constellation point, while the output represents the amplitude and angle of the transformed constellation point (the conversion from the amplitude-angle-domain to complex domain is trivial and is omitted from the notation for brevity). The neural network model for producing the weight factors, denoted by $W_{\bm{\theta},l,c}(\mathbf{q}_\textrm{QAM})$, follows a similar architecture, i.e., it is a fully-connected network with four hidden layers. However, the number of neurons is somewhat smaller, ranging from 8 to 16 per layer, while the activation functions of the hidden layers are rectified linear units. Moreover, the output layer uses the \emph{softmax} activation function to output $C$ values which have a sum of $1$. This makes them suitable to be used as weighting factors for the linear combination. In our experiments, we observed best performance with $C=3$ and used this value in all the results.

\subsection{Pilotless DeepRx Receiver and End-to-End Training}

The input of the DeepRx is formed by collecting the received signal samples over one slot, consisting of $N_F$ subcarriers and $N_S$ OFDM symbols. Therefore, the input array can be written as $\mathbf{Y} \in \mathcal{C}^{N_F \times N_T \times N_R}$. As mentioned above, this input is fed to a MIMO DeepRx model utilizing a fully learned multiplicative transformation, as described in \cite{Korpi21}. However, now there is no separate input for the raw channel estimate as there are no pilots being transmitted. Moreover, the model size is somewhat larger, with the number of convolutional filters for each ResNet block ranging from 512 to 2048. The output of the pilotless MIMO DeepRx model consists again of the log-likelihood ratios (LLRs), which are fed to the low-density parity check (LDPC) decoder for extracting the information bits.

The system can be trained end-to-end assuming that we have a differentiable implementation of the radio link (e.g., using \cite{sionna}). In this work, the training is done similar to \cite{Honkala21} by using the binary cross entropy (BCE) between the detected and transmitted bits as the loss function. With this, the BCE loss term can be written as follows:
\begin{align}
\mathrm{BCE}_q \left(\bm{\theta} \right) = -\frac{1}{W_q} \sum_{i=0}^{W_q - 1} & \left( b_{iq} \operatorname{log}\left(\hat{b}_{iq}\right) \right.\nonumber\\
&+ \left. \left(1-b_{iq}\right) \operatorname{log}\left(1-\hat{b}_{iq} \right) \right),
\end{align}
where $q$ is the sample index within the batch, $b_{iq}$ is the transmitted bit, $\hat{b}_{iq}$ is the bit probability estimate at the DeepRx output (after applying the Sigmoid function to convert LLRs to probabilities \cite{Honkala21}), and $W_q$ is the total number of transmitted bits within the slot.

In addition to the BCE loss, an additional term is introduced to improve the convergence of the training process with larger constellation sizes. In particular, it turns out that the end-to-end trained system often ends up trying to transmit only some of the bits within each symbol. This is due to tendency to learn clusters of constellation points such that points in each cluster are very tightly spaced. Such a constellation is a local minima from which it is often difficult to escape.  For instance, if the transmitter learns to divide the constellation points only into four clusters, it can transmit 2 bits per symbol with high accuracy but whenever it tries to split the clusters to transmit more bits, the accuracy will temporarily decrease. To remove the possibility of such a naive solution, we introduce a loss term which penalizes for very closely spaced constellation points as follows:
\begin{align}
D_q \left(\bm{\theta} \right) = \operatorname{ReLu} \left(\operatorname{log}\left(\frac{1}{N_T} \sum_{l=0}^{N_T}\left(\frac{d_{l,\text{max}}\left(\bm{\theta}\right)}{d_{l,\text{min}}\left(\bm{\theta}\right)} \right)\right) - b \right)
\end{align}
where $d_{l,\text{max}}\left(\bm{\theta}\right)$ and $d_{l,\text{min}}\left(\bm{\theta}\right)$ are the maximum and minimum distances between two constellation points for the $l$th layer, respectively, $b$ is a predefined bias term, and ReLu is the rectified linear unit activation function. The purpose of the ReLu function, together with the bias $b$, is to render all sufficiently small loss terms to zero, which effectively means that this loss term impacts the training only at the beginning phase. In our work, a value of $b=4.5$ is used with 16-point constellations, and $b=7.5$ with 64-point constellations.%\todo{$b$:n sijaan oli $B$, meniko oikein?}

The overall loss term for a single sample is
\begin{align}
L_q \left(\bm{\theta} \right) = \operatorname{log}\left(1+ \mathrm{snr}\right) \mathrm{BCE}_q \left(\bm{\theta} \right) + \lambda D_q \left(\bm{\theta} \right),
\end{align}
where $\mathrm{snr}$ is the signal-to-noise ratio (SNR) on a linear scale and $\lambda$ is the weight factor of the distance-based loss term. A value of $\lambda$ is experimentally chosen to be $\lambda=0.1$ for 16-point constellations and $\lambda=0.05$ for 64-point constellations. The training is done using the ADAM optimizer, with a batch size of $10$ and learning rate of $5\cdot 10^{-4}$. When the transmitter and DeepRx are trained jointly using the above loss function, DeepRx will learn to utilize the simultaneously learned constellations for detecting the data signals without any pilots.

\section{Simulation Results}

\begin{table}
  \setlength{\tabcolsep}{3pt}
    \renewcommand{\arraystretch}{1.3}
    \footnotesize
    \centering
    \caption{Simulation parameters for training and validation. The parameters with more than one value are randomized using a uniform distribution.}
    \begin{tabular}{|l|p{2.1cm}|p{1.2cm}|}
    \hline
    \multicolumn{1}{|c|}{\textbf{Parameter}} & \textbf{Training} & \textbf{Validation}\\
		\hline\hline
		Carrier frequency & \multicolumn{2}{c|}{3.5 GHz} \\
    \hline
		Channel model & CDL-A, CDL-B & CDL-C \\
		\hline
		Signal-to-noise ratio (SNR) & \multicolumn{2}{c|}{0--30 dB}\\
		\hline
		RMS delay spread & \multicolumn{2}{c|}{10 ns -- 300 ns} \\
		\hline
		Velocity & \multicolumn{2}{c|}{0 m/s -- 5 m/s} \\
		\hline
		Number of subcarriers & \multicolumn{2}{c|}{72} \\
		\hline
		Subcarrier spacing & \multicolumn{2}{c|}{30 kHz} \\
		\hline
		Slot length & \multicolumn{2}{c|}{14 OFDM symbols} \\
		\hline
		MCS & \multicolumn{2}{c|}{See Table~\ref{table:mcs}} \\
		\hline
		Number of RX antennas & \multicolumn{2}{c|}{4} \\
		\hline
		Number of TX streams & \multicolumn{2}{c|}{2} \\
		\hline
		Pilot configuration (DMRS baseline only) & \multicolumn{2}{c|}{2 DMRSs per slot} \\
		\hline
    \end{tabular}
    \label{table:param}
  \end{table}

\begin{table}
  \setlength{\tabcolsep}{3pt}
    \renewcommand{\arraystretch}{1.3}
    \footnotesize
    \centering
    \caption{Considered MCS values, representing reasonable modulation order and code rate combinations for 16- and 64-point constellations.}
    \begin{tabular}{|l|c|c|}
    \hline
    \textbf{MCS index} & \textbf{Bits per symbol} & \textbf{Code rate}\\
		\hline
		1 & 4 (16-QAM) & 0.37 \\
    \hline
		2 & 4 (16-QAM) & 0.42 \\
    \hline
		3 & 4 (16-QAM) & 0.48 \\
    \hline
		4 & 4 (16-QAM) & 0.54 \\
    \hline
		5 & 4 (16-QAM) & 0.60 \\
    \hline
		6 & 4 (16-QAM) & 0.64 \\
    \hline
		7 & 6 (64-QAM) & 0.46 \\
    \hline
		8 & 6 (64-QAM) & 0.50 \\
    \hline
		9 & 6 (64-QAM) & 0.55 \\
    \hline
		10 & 6 (64-QAM) & 0.60 \\
    \hline
		11 & 6 (64-QAM) & 0.65 \\
    \hline
		12 & 6 (64-QAM) & 0.70 \\
    \hline
		13 & 6 (64-QAM) & 0.75 \\
    \hline
		14 & 6 (64-QAM) & 0.80 \\
    \hline
		15 & 6 (64-QAM) & 0.85 \\
    \hline
    \end{tabular}
    \label{table:mcs}
  \end{table}

%/mnt/shared/korpi/checkpoints_new/mimo_run_test_v1_mn35_c3
\newcommand{\BLERQMFOUR}{plot_data_v2/qm4_0_42/valid_blers.csv}
\newcommand{\constlayeronefour}{plot_data_v2/qm4_0_42/valid_layer0_const.csv}
\newcommand{\constlayertwofour}{plot_data_v2/qm4_0_42/valid_layer1_const.csv}

%/mnt/shared/korpi/checkpoints_new/mimo_run_test_v2_qm6_mn35_a2
\newcommand{\BLERQMSIX}{plot_data_v2/qm6_0_50/valid_blers.csv}
\newcommand{\constlayeronesix}{plot_data_v2/qm6_0_50/valid_layer0_const.csv}
\newcommand{\constlayertwosix}{plot_data_v2/qm6_0_50/valid_layer1_const.csv}

\newcommand{\mcsdata}{plot_data_v2/LA/link_adapt_Oct11_scaled.csv}
\newcommand{\mcsindml}{plot_data_v2/LA/mcs_info_ml.csv}
\newcommand{\mcsindperf}{plot_data_v2/LA/mcs_info_perf.csv}
\newcommand{\mcsindprac}{plot_data_v2/LA/mcs_info_prac.csv}

\newcommand{\segain}{plot_data_v2/LA/gain_Oct11.csv}

\begin{figure}[!t]
\begin{tikzpicture}
\begin{axis}[
width=0.99\columnwidth,
height=0.8\columnwidth,
ymode=log,
xmin=0,
xmax=14,
ymin=2e-3,
ymax=1,
grid=both,
xlabel={SNR (dB)},
ylabel={BLER},
legend cell align={left},
legend pos=south west,
legend style={font=\scriptsize},
font = \small
]
\addplot[mark=o, mark size=3pt, solid,red,line width=1pt] table[x=SNR, y=practical, col sep=comma] {\BLERQMFOUR};
\addlegendentry{K-Best with DMRS}
\addplot[mark=square*, mark size=2pt, mark options={solid, fill=blue}, dashed,blue,line width=1pt] table[x=SNR, y=perfect, col sep=comma] {\BLERQMFOUR};
\addlegendentry{K-Best with perfect CSI}
\addplot[mark=diamond, mark size=3pt, solid,black,line width=1pt] table[x=SNR, y=ML, col sep=comma] {\BLERQMFOUR};
\addlegendentry{ML-based pilotless scheme}
\end{axis}
\end{tikzpicture}
\caption{BLER performance of the different solutions with 4 bits per symbol (MCS index 2).}
\label{fig:basic_bler_qm4}
\end{figure}
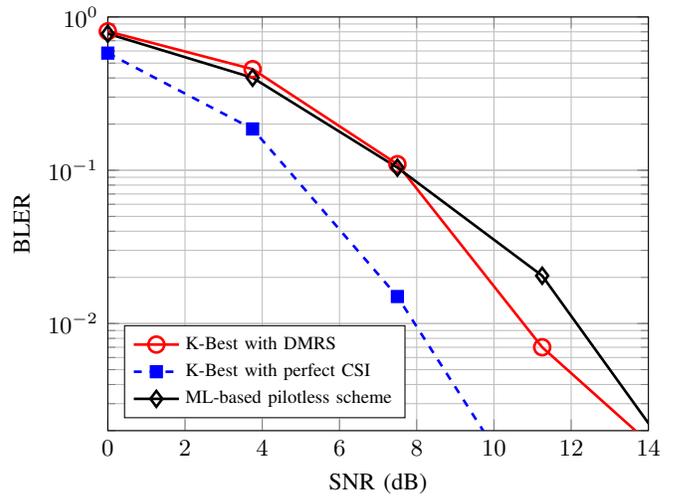

\begin{figure}[!t]
\begin{tikzpicture}
\begin{axis}[
width=0.99\columnwidth,
height=0.8\columnwidth,
ymode=log,
xmin=0,
xmax=25,
ymin=2e-3,
ymax=1,
grid=both,
xlabel={SNR (dB)},
ylabel={BLER},
legend cell align={left},
legend pos=south west,
legend style={font=\scriptsize},
font = \small
]
\addplot[mark=o, mark size=3pt, solid,red,line width=1pt] table[x=SNR, y=practical, col sep=comma] {\BLERQMSIX};
\addlegendentry{K-Best with DMRS}
\addplot[mark=square*, mark size=2pt, mark options={solid, fill=blue}, dashed,blue,line width=1pt] table[x=SNR, y=perfect, col sep=comma] {\BLERQMSIX};
\addlegendentry{K-Best with perfect CSI}
\addplot[mark=diamond, mark size=3pt, solid,black,line width=1pt] table[x=SNR, y=ML, col sep=comma] {\BLERQMSIX};
\addlegendentry{ML-based pilotless scheme}
\end{axis}
\end{tikzpicture}
\caption{BLER performance of the different solutions with 6 bits per symbol (MCS index 8).}
\label{fig:basic_bler_qm6}
\end{figure}
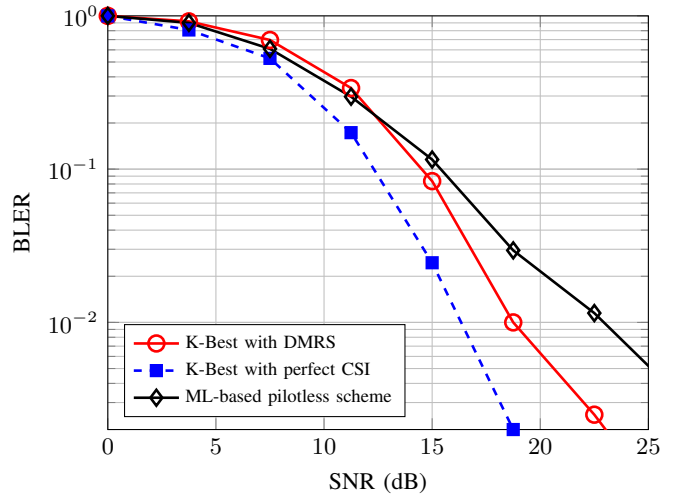

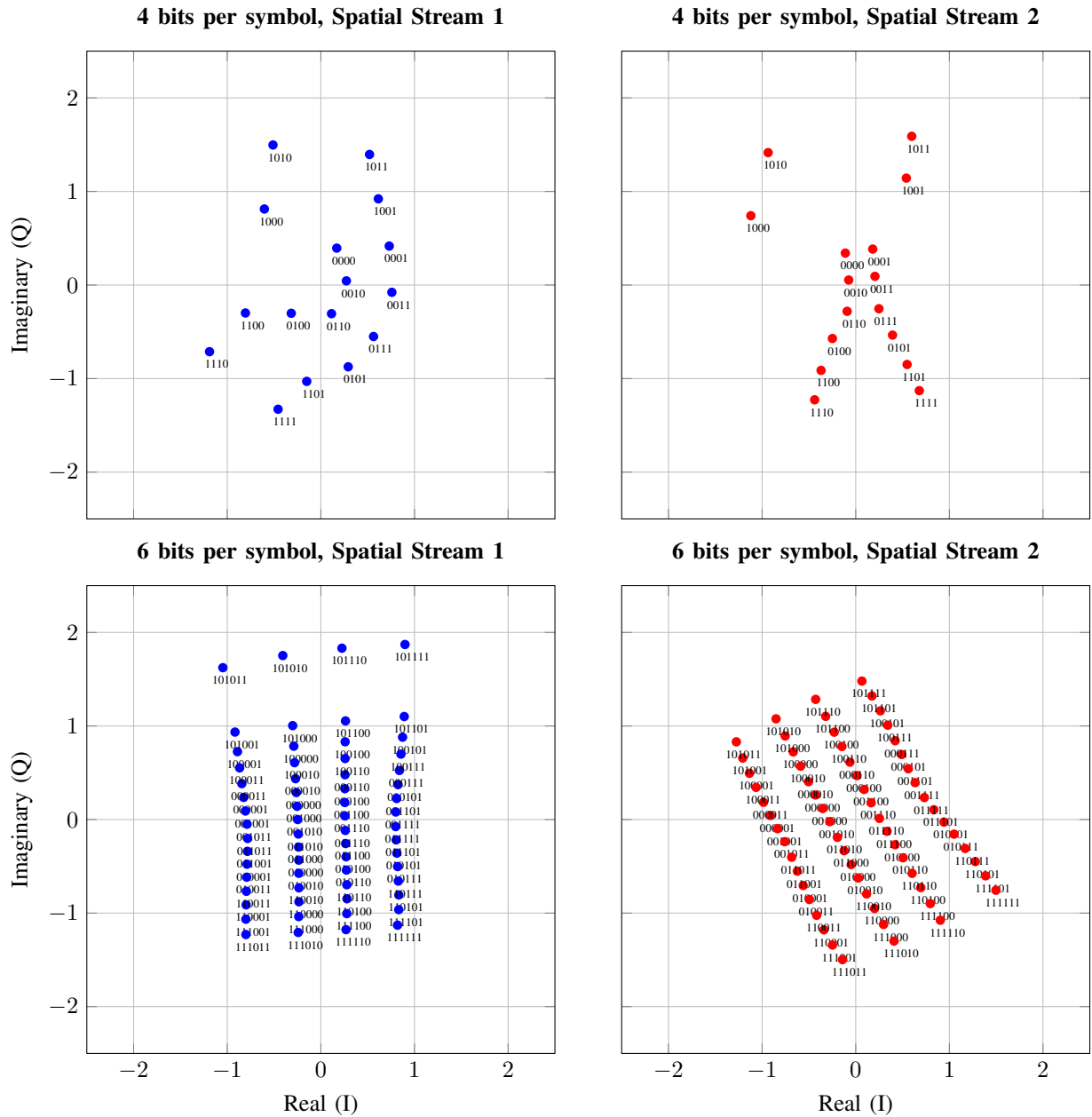
\begin{figure*}[ht!]
\centering
\begin{tikzpicture}
\begin{groupplot}[
group style={group size=2 by 2,
xlabels at=edge bottom,
xticklabels at=edge bottom,
ylabels at=edge left,
yticklabels at=edge left,},
height=0.97\columnwidth,
width=0.97\columnwidth,
xmin=-2.5,
xmax=2.5,
ymin=-2.5,
ymax=2.5,
grid=both,
xlabel={Real (I)},
ylabel={Imaginary (Q)}]
\nextgroupplot[title=\textbf{4 bits per symbol, Spatial Stream 1}]
\addplot[
		scatter,
    blue,
		only marks,
    mark=*,
    mark options={fill=blue, draw opacity=0},
    nodes near coords, % Place nodes near each coordinate
    point meta=explicit symbolic, % The meta data used in the nodes is not explicitly provided and not numeric
    every node near coord/.style={anchor=120, font=\tiny, color=black} % Align each coordinate at the anchor 40 degrees clockwise from the right edge
    ] table [meta index=2,x=x, y=y, col sep=comma] {\constlayeronefour};
\nextgroupplot[title=\textbf{4 bits per symbol, Spatial Stream 2}]
\addplot[
		scatter,
    red,
		only marks,
    mark=*,
    mark options={fill=red, draw opacity=0},
    nodes near coords, % Place nodes near each coordinate
    point meta=explicit symbolic, % The meta data used in the nodes is not explicitly provided and not numeric
    every node near coord/.style={anchor=120, font=\tiny, color=black} % Align each coordinate at the anchor 40 degrees clockwise from the right edge
    ] table [meta index=2,x=x, y=y, col sep=comma] {\constlayertwofour};
\nextgroupplot[title=\textbf{6 bits per symbol, Spatial Stream 1}]
\addplot[
		scatter,
    blue,
		only marks,
    mark=*,
    mark options={fill=blue, draw opacity=0},
    nodes near coords, % Place nodes near each coordinate
    point meta=explicit symbolic, % The meta data used in the nodes is not explicitly provided and not numeric
    every node near coord/.style={anchor=120, font=\tiny, color=black} % Align each coordinate at the anchor 40 degrees clockwise from the right edge
    ] table [meta index=2,x=x, y=y, col sep=comma] {\constlayeronesix};
	\nextgroupplot[title=\textbf{6 bits per symbol, Spatial Stream 2}]
\addplot[
		scatter,
    red,
		only marks,
    mark=*,
    mark options={fill=red, draw opacity=0},
    nodes near coords, % Place nodes near each coordinate
    point meta=explicit symbolic, % The meta data used in the nodes is not explicitly provided and not numeric
    every node near coord/.style={anchor=120, font=\tiny, color=black} % Align each coordinate at the anchor 40 degrees clockwise from the right edge
    ] table [meta index=2,x=x, y=y, col sep=comma] {\constlayertwosix};
\end{groupplot}
\end{tikzpicture}
\caption{Learned constellation for both spatial streams and both modulation orders.}
\label{fig:const_all}
\vspace{-2mm}
\end{figure*}

\begin{figure}[!t]
\begin{tikzpicture}
\begin{axis}[
width=\columnwidth,
height=0.8\columnwidth,
xmin=7,
xmax=20,
ymin=2,
ymax=9,
grid=both,
xlabel={SNR (dB)},
ylabel={Spectral efficiency (bps/Hz)},
legend cell align={left},
legend pos=south east,
legend style={font=\scriptsize},
font = \small
]
\addplot[solid,red,line width=1pt] table[x=SNR, y=practical, col sep=comma] {\mcsdata};
\addlegendentry{K-Best with DMRS}
\addplot[blue,line width=1pt] table[x=SNR, y=perfect, col sep=comma] {\mcsdata};
\addlegendentry{K-Best with perfect CSI}
\addplot[solid,black,line width=1pt] table[x=SNR, y=ML, col sep=comma] {\mcsdata};
\addlegendentry{ML-based pilotless}
\addplot[
		scatter,
		only marks,
    mark=,
    nodes near coords, % Place nodes near each coordinate
    point meta=explicit symbolic, % The meta data used in the nodes is not explicitly provided and not numeric
    every node near coord/.style={anchor=-180, font=\scriptsize, color=red} % Align each coordinate at the anchor 40 degrees clockwise from the right edge
    ] table [meta index=2,x=x, y=y, col sep=comma] {\mcsindprac};
\addplot[
		scatter,
		only marks,
    mark=,
    nodes near coords, % Place nodes near each coordinate
    point meta=explicit symbolic, % The meta data used in the nodes is not explicitly provided and not numeric
    every node near coord/.style={anchor=-180, font=\scriptsize, color=blue} % Align each coordinate at the anchor 40 degrees clockwise from the right edge
    ] table [meta index=2,x=x, y=y, col sep=comma] {\mcsindperf};
\addplot[
		scatter,
		only marks,
    mark=,
    nodes near coords, % Place nodes near each coordinate
    point meta=explicit symbolic, % The meta data used in the nodes is not explicitly provided and not numeric
    every node near coord/.style={anchor=-180, font=\scriptsize, color=black} % Align each coordinate at the anchor 40 degrees clockwise from the right edge
    ] table [meta index=2,x=x, y=y, col sep=comma] {\mcsindml};
\end{axis}
\end{tikzpicture}
\caption{The spectral efficiency of the considered schemes with BLER of 10\%. The used MCS indices are also shown.}
\label{fig:se_plot}
\end{figure}
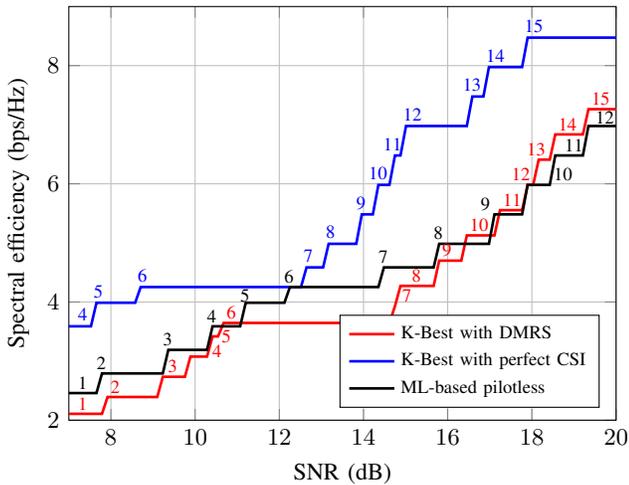

% NOTE: You need to specify manually the segments in which the gain is negative (indexing starts from 0).
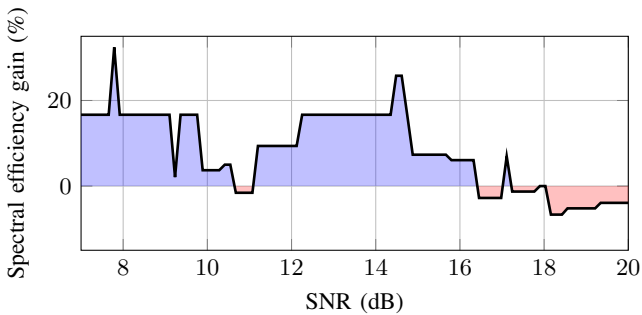
\begin{figure}[!t]
\begin{tikzpicture}
\begin{axis}[
width=\columnwidth,
height=0.5\columnwidth,
xmin=7,
xmax=20,
ymin=-15,
ymax=35,
grid=both,
xlabel={SNR (dB)},
ylabel={Spectral efficiency gain (\%)},
legend cell align={left},
legend pos=south east,
legend style={font=\scriptsize},
font = \small
]
\addplot[name path=f, black,line width=1pt] table[x=SNR, y=gain, col sep=comma] {\segain};
\path[name path=axis] (axis cs:7,0) -- (axis cs:20,0);
\addplot [thick, color=blue, fill=blue, fill opacity=0.25]
fill between[of=f and axis,split,every segment no 1/.style={red},every segment no 3/.style={red},every segment no 5/.style={red},every segment no 6/.style={red}];
\end{axis}
\end{tikzpicture}
\caption{The spectral efficiency gain of the ML-based pilotless scheme over the conventional pilot-based scheme.}
\label{fig:se_gain}
\end{figure}

In order to demonstrate the effectiveness of the proposed approach in improving spectral efficiency, single-user MIMO (SU-MIMO) simulations were performed using the Sionna library \cite{sionna}. The simulation parameters are presented in Table~\ref{table:param}, while the considered modulation and coding schemes (MCSs), chosen similar to 5G NR MCS index table 2 \cite{NR_38214}, are shown in Table~\ref{table:mcs}. In particular, the proposed approach is trained with both 16 and 64 point constellations, resulting in 4 and 6 bits per symbol, respectively. Different channels models are employed for the training and validation to avoid overfitting to the selected channel models. For training, each training sample is generated either from CDL-A and CDL-B channel profiles (selected randomly for each training sample), while CDL-C is used for validation \cite{NR_38901}. The proposed learned pilotless scheme is compared to a conventional QAM-OFDM waveform and a nonlinear K-Best detector using either perfect channel state information (CSI) or a demodulation reference signal (DMRS) based channel estimate. The former represents essentially the highest achievable performance under the considered simulation scenario, while the latter utilizes a DMRS pattern consisting of two pilot-carrying OFDM symbols within the slot, with the pilots located on the 3rd and 12th OFDM symbols.

Considering first the block error rate (BLER) achieved with 4 bits per symbol (MCS index 2), the results are shown in Fig.~\ref{fig:basic_bler_qm4}. It can be observed that the proposed pilotless scheme essentially matches the practical baseline when the BLER is in the order of 10\%, despite resorting to completely blind detection at the receiver side. Since the pilotless scheme can use all resource elements (REs) for data transmission, unlike the DMRS-based scheme, this of course translates to a spectral efficiency gain. The same holds also for the case where there are 6 bits per symbol (MCS index 8), whose BLER results are shown in Fig.~\ref{fig:basic_bler_qm6}. Again, the proposed learned pilotless scheme nearly matches the performance of the practical pilot-based receiver under the relevant BLER values.

To provide some insight into how the ML-based scheme achieves accurate pilotless detection, let us observe the learned constellations in Fig.~\ref{fig:const_all}. It is evident that the learned constellations are asymmetric, which is an obvious requirement for blind and pilotless detection. Moreover, when considering the differences in the constellations of the overlapping spatial streams, it can be observed that at least one of the overlapping constellations has some outliers. This is likely an useful feature for the considered DeepRx receiver when it learns to do pilotless separation of the spatial streams, as such differing outliers are more easily detectable. Investigating the necessary requirements of the constellation shapes further is an important future work item for us.

In order to compare the different schemes in terms of the achieved throughput, Fig.~\ref{fig:se_plot} shows their spectral efficiencies with respect to the SNR. The spectral efficiency is obtained by selecting the highest MCS index from Table~\ref{table:mcs} with which a BLER target of 10\% is achieved with each scheme, and then calculating what is the actual throughput that is achieved with this MCS, taking into account also the pilot overhead. In addition, also the MCS index utilized by each scheme is indicated on the plot. It can be observed that at the lower SNRs the ML-based pilotless scheme achieves the BLER target with the same MCS as the pilot-based scheme, which translates to a higher spectral efficiency. However, when the 64-point constellation is used (from MCS index 7 upwards), the spectral efficiency of the ML-based scheme starts to deteriorate compared to the conventional DMRS-based system. Indeed, when the SNR goes above 16~dB, the spectral efficiency of the ML-based scheme is slightly lower than that of the DMRS-based scheme.

This phenomenon can be analyzed more conveniently from Fig.~\ref{fig:se_gain} which shows the spectral efficiency gain of the ML-based scheme over the conventional DMRS-based scheme. With SNRs below 15 dB, it achieves typically a gain of 15-20\%, with some exceptions. However, when the SNR goes above 15~dB, the performance gain is almost completely lost. Since use of the 64-point constellations is required at these SNRs, the issue is most likely related to the higher-order constellation. It is likely that a larger DeepRx model would be required for detecting the high-order modulation symbols at high SNRs. Another alternative could be to learn SNR-dependent constellation shapes, considering that the current scheme uses a fixed constellation regardless of the SNR.

\section{Conclusion}

This paper proposed an ML-based approach for learning pilotless spatial multiplexing. In particular, it was shown that by learning jointly the constellation shapes used by individual spatial streams as well as a convolutional neural network-based receiver, it is possible to transmit several spatial MIMO streams successfully without any pilots. This results in as much as a 15\% to 20\% increase in spectral efficiency compared to a conventional pilot-based MIMO system. As future work, we plan to study in more detail the requirements for the constellation shapes, as well as address the performance challenges at high signal-to-noise ratios.

\section*{Acknowledgments}

This work has been partly funded by the European Commission through the project Hexa-X-II (Grant Agreement no. 101095759).

\bibliographystyle{IEEEtran}
% argument is your BibTeX string definitions and bibliography database(s)
\bibliography{IEEEabrv,references}

% Generated by IEEEtran.bst, version: 1.12 (2007/01/11)
\begin{thebibliography}{10}
\providecommand{\url}[1]{#1}
\csname url@samestyle\endcsname
\providecommand{\newblock}{\relax}
\providecommand{\bibinfo}[2]{#2}
\providecommand{\BIBentrySTDinterwordspacing}{\spaceskip=0pt\relax}
\providecommand{\BIBentryALTinterwordstretchfactor}{4}
\providecommand{\BIBentryALTinterwordspacing}{\spaceskip=\fontdimen2\font plus
\BIBentryALTinterwordstretchfactor\fontdimen3\font minus
  \fontdimen4\font\relax}
\providecommand{\BIBforeignlanguage}[2]{{%
\expandafter\ifx\csname l@#1\endcsname\relax
\typeout{** WARNING: IEEEtran.bst: No hyphenation pattern has been}%
\typeout{** loaded for the language `#1'. Using the pattern for}%
\typeout{** the default language instead.}%
\else
\language=\csname l@#1\endcsname
\fi
#2}}
\providecommand{\BIBdecl}{\relax}
\BIBdecl

\bibitem{Honkala21}
M.~Honkala, D.~Korpi, and J.~M.~J. Huttunen, ``{DeepRx}: Fully convolutional
  deep learning receiver,'' \emph{IEEE Transactions on Wireless
  Communications}, vol.~20, no.~6, pp. 3925--3940, Jun. 2021.

\bibitem{Hoydis21a}
J.~Hoydis, F.~A. Aoudia, A.~Valcarce, and H.~Viswanathan, ``Toward a {6G}
  {AI}-native air interface,'' \emph{IEEE Communications Magazine}, vol.~59,
  no.~5, pp. 76--81, 2021.

\bibitem{aoudia20}
F.~A. Aoudia and J.~Hoydis, ``End-to-end learning for {OFDM}: From neural
  receivers to pilotless communication,'' \emph{IEEE Transactions on Wireless
  Communications}, vol.~21, no.~2, pp. 1049--1063, Feb. 2022.

\bibitem{He20}
H.~He, C.-K. Wen, S.~Jin, and G.~Y. Li, ``Model-driven deep learning for {MIMO}
  detection,'' \emph{IEEE Transactions on Signal Processing}, vol.~68, pp.
  1702--1715, Feb. 2020.

\bibitem{Korpi21}
D.~Korpi, J.~Huttunen, and M.~Honkala, ``{DeepRx} {MIMO}: Convolutional {MIMO}
  detection with learned multiplicative transformations,'' in \emph{Proc. IEEE
  International Conference on Communications (ICC)}, Jun. 2021.

\bibitem{pratik20}
K.~Pratik, B.~Rao, and M.~Welling, ``{RE-MIMO}: Recurrent and permutation
  equivariant neural {MIMO} detection,'' \emph{IEEE Transactions on Signal
  Processing}, vol.~69, pp. 459--473, Dec. 2020.

\bibitem{Huang18}
Y.-D. Huang, P.~P. Liang, Q.~Zhang, and Y.-C. Liang, ``A machine learning
  approach to {MIMO} communications,'' in \emph{Proc. IEEE International
  Conference on Communications (ICC)}, May 2018.

\bibitem{Liao20}
J.~Liao, J.~Zhao, F.~Gao, and G.~Y. Li, ``A model-driven deep learning method
  for massive {MIMO} detection,'' \emph{IEEE Communications Letters}, vol.~24,
  no.~8, pp. 1724--1728, Apr. 2020.

\bibitem{sionna}
J.~Hoydis, S.~Cammerer, F.~{Ait Aoudia}, A.~Vem, N.~Binder, G.~Marcus, and
  A.~Keller, ``Sionna: An open-source library for next-generation physical
  layer research,'' \emph{arXiv preprint}, Mar. 2022.

\bibitem{NR_38214}
``{Technical Specification Group Radio Access Network; NR; Physical layer
  procedures for data ({3GPP TS} 38.214 version 18.0.0 Release 18)},'' ETSI,
  Sophia Antipolis Cedex, France, Sep. 2023.

\bibitem{NR_38901}
``{3rd Generation Partnership Project; Technical Specification Group Radio
  Access Network; Study on channel model for frequencies from 0.5 to 100 {GHz}
  ({3GPP TR} 38.901 version 17.0.0 Release 17)},'' ETSI, Sophia Antipolis
  Cedex, France, Mar. 2022.

\end{thebibliography}

\end{document}